\documentclass[12pt]{article}

\usepackage{amssymb}
\usepackage{amsmath}
\usepackage{amsfonts}
\usepackage{graphicx}
\usepackage{lineno,hyperref}
\usepackage[english]{babel}
\usepackage{tabularx}
\usepackage{subcaption}
\usepackage{xcolor}
\usepackage{authblk}

\usepackage[round, sort]{natbib}
\newcommand{\mb}{\mathbf}
\newcommand{\mbb}{\mathbb}
\newcommand{\bit}{\begin{itemize}}
\newcommand{\eit}{\end{itemize}}
\newcommand{\ben}{\begin{enumerate}}
\newcommand{\een}{\end{enumerate}}
\newcommand{\bc}{\end{center}}
\newcommand{\ec}{\end{center}}
\newcommand{\beq}{\begin{equation}}
\newcommand{\bsb}{\boldsymbol}
\newcommand{\eeq}{\end{equation}} 

\begin{document}

\title{A constrained ICA-EMD Model for Group Level fMRI Analysis}

\author[1,3]{S. Wein \thanks{Simon.Wein$@$psychologie.uni-regensburg.de}}
\author[2]{A.~M.~Tom\'e}
\author[1,3]{M. Goldhacker}
\author[3]{M. W. Greenlee}
\author[1]{E. W. Lang}

\affil[1]{CIML, Biophysics, University of Regensburg, 93040 Regensburg, Germany}
\affil[2]{IEETA/DETI, Universidade de Aveiro, 3810-193 Aveiro, Portugal}
\affil[3]{Experimental Psychology, University of Regensburg, 93040 Regensburg, Germany}

\maketitle

%%%%%%%%%%%%%%%%%%%%%%%%%%%%%%%%%%
%%%%%%%%%%%%%%%%%%%%%%%%%%%%%%%%%%
%%%%%%%%%%%%%%%%%%%%%%%%%%%%%%%%%%

\begin{abstract}
Independent component analysis (ICA), as a data driven method, has shown to be a powerful tool for functional magnetic resonance imaging (fMRI) data analysis. One drawback of this multivariate approach is, that it is not compatible to the analysis of group data in general. Therefore various techniques have been proposed in order to overcome this limitation of ICA. In this paper a novel ICA-based work-flow for extracting resting state networks from fMRI group studies is proposed. An empirical mode decomposition (EMD) is used to generate reference signals in a data driven manner, which can be incorporated into a constrained version of ICA (cICA), what helps to eliminate the inherent ambiguities of ICA. The results of the proposed workflow are then compared to those obtained by a widely used group ICA approach for fMRI analysis. In this paper it is demonstrated that intrinsic modes, extracted by EMD, are suitable to serve as references for cICA to obtain typical resting state patterns, which are consistent over subjects. By introducing these reference signals into the ICA, our processing pipeline makes it transparent for the user, how comparable activity patterns across subjects emerge. This additionally allows adapting the trade-off between enforcing similarity across subjects and preserving individual subject features.
\end{abstract}

%%%%%%%%%%%%%%%%%%%%%%%%%%%
%%%%%%%%%%%%%%%%%%%%%%%%%%%
%%%%%%%%%%%%%%%%%%%%%%%%%%% 
\section{Introduction}

Independent component analysis (ICA) is a data driven tool, which is widely employed for functional magnetic resonance imaging (fMRI) data analysis. Based on a linear mixture model, either spatially \citep{McKeown1998} or temporally \citep{Biswal1999} independent components (ICs) can be obtained with ICA, without the requirement of prior information in form of anatomical regions of interest or temporal activation profiles. One problem of ICA is that, because of inherent indeterminacies, it is in general not suitable for group studies. Different subjects have different time courses and spatial maps, and the extracted components will be sorted differently. This can make it difficult to find comparable activation patterns between subjects and draw inferences from subject groups. So far various approaches have been proposed in order to overcome these shortcomings of ICA \citep{Calhoun2009}. Combining components obtained by single subject ICA-based on spatial correlation or clustering was proposed by \cite{Calhoun2001} and \cite{Esposito2005}. Another possibility is the spatial or temporal concatenation of the individual datasets to obtain components in a single ICA step from a group dataset and the employment of back-reconstruction approaches to obtain subject specific components \citep{Svensn2002, Calhoun2001_2}. These concatenation-based approaches were compared with a simple across-subject averaging by \cite{Schmithorst2004}. A more involved approach considered a tensorial extension of ICA, presented by \cite{Beckmann2005}. The authors extended the probabilistic ICA (PICA) model by adding a third dimension representing subject-related dependencies in addition to the spatio-temporal dimensions. The model represents a three-way factor analysis similar to the well-known PARAFAC model \citep{Harshman1994}. 

A popular paradigm used to generate data for applying the above mentioned exploratory matrix factorization techniques is the so-called resting-state. During this paradigm, subjects either rest with eyes open fixating a fixation cross, or with eyes closed. Usually, subjects are instructed not to fall asleep and to let their mind wander. Contrary to the simplicity of the paradigm, the generated database has complex spatial structure and temporal dynamics, which arise from low-frequency fluctuations in the BOLD signal \citep{Fox2007}. Furthermore the data is characterized by large spatial dimension and the lack of triggers or any underlying paradigm, which are usually used in task-based fMRI investigations. Because of the latter two aspects, exploratory matrix factorization techniques are appropriate to handle the large amount of data and explore the complex spatial and temporal structure. In this context, ICA-based pipelines have emerged as a state-of-the-art approach to investigate rs-fMRI data \citep{Allen2011,Remes2011}. This decomposition of resting-state fMRI data results in a so-called parcellation of the cortex into brain networks composed of functionally similar brain areas. In the literature, common brain networks are default-mode, cognitive control, visual, somatomotor, sub-cortical, auditory, cerebellar, depending on the function of the brain areas included in each network \citep{Allen2012}.

In this paper, a hybrid method is proposed for extracting resting state networks (RSNs) from fMRI data, based on constrained ICA (cICA) and empirical mode decomposition (EMD). This constrained extension of ICA optimizes besides the statistical independence, the similarity to a given reference signal. Incorporating a reference into ICA, in the framework of an augmented Lagrangian approach, helps to obtain more robust ICs and can eliminate the ambiguities of ICA \citep{Lu2006, Lin2009, Rodriguez2014}. In this paper, like \citep{Lin2009}, spatial reference maps were employed to extract resting state networks from fMRI data. It is shown that an EMD based image decomposition technique, denoted as Green's function in tension based bi-dimensional ensemble EMD (GiT-BEEMD) \citep{Albaddai2016}, produces suitable references for cICA. This two-dimensional extension of EMD allows to decompose images into so called bi-dimensional intrinsic mode functions (BIMFs) and can also be used to decompose volumetric fMRI images slice-wise. Because of its inherent natural ordering of the extracted intrinsic modes according to their spatial frequencies, EMD can easily generate prototype spatial maps. Similar spatial maps obtained with the EMD for each subject can be identified and averaged across subjects. In a next step these prototype spatial maps can serve as reference signals for a constrained ICA applied in parallel to the entire group of subjects. In this workflow the references are obtained from the same dataset as used for the analysis, so no prior information is required. A resting state fMRI dataset from the \textit{Human Connectome Project} \citep{Vanessen2012} is used to compare the results to those obtained by the widely used group ICA (gICA) based on temporal concatenation \citep{Calhoun2001_2}. Finally potential benefits of the proposed approach are emphasized, by showing that this approach allows the user to actively shape the extracted resting state networks. The trade-off between enforcing a certain similarity across subjects and preserving individual subject features can be determined, and can be well adapted to optimally fulfill the requirements of different studies. 

%%%%%%%%%%%%%%%%%%%%%%%%%%%
%%%%%%%%%%%%%%%%%%%%%%%%%%%
%%%%%%%%%%%%%%%%%%%%%%%%%%%

\section{Material and Methods}

The following subsections introduce the dataset employed, and describe the data analysis techniques, which combine cICA and GiT-BEEMD, as well as the processing steps of the gICA approach used for comparison. Also a flowchart of the proposed signal processing chain is provided. 

%%%%%%%%%%%%%%%%%%%%%%%%%%%
%%%%%%%%%%%%%%%%%%%%%%%%%%%
\subsection{Dataset}

For this study a data set from the \textit{Human Connectome Project} was employed \citep{Vanessen2012}. The S1200 release includes data from subjects which participated in four resting state fMRI sessions, lasting 14.4 minutes each and resulting in 1200 volumes per session. Customized Siemens \textit{Connectome Skyra} magnetic resonance imaging (MRI) scanners at Washington University with a field strength of $\bf{B}_0 = 3\ Tesla$ were employed for data acquisition, using a multi-band (factor 8) technique \citep{Moeller2010, Feinberg2010, Setsompop2012, Xu2012}. The data was collected by gradient-echo echo-planar imaging (EPI) sequences with a repetition time $TR = 720\ ms$ and an echo time $TE = 31.1\ ms$, using a flip angle of $\theta = 52^{\circ}$. The field of view was $FOV = 208\ mm \times 180\ mm$ and $N_s = 72$ slices with a thickness of $d_s = 2\ mm$ were obtained, containing voxels with a size of $2\ mm \times  2\ mm \times 2\ mm$. The preprocessed version, including motion-correction, structural preprocessing and ICA-FIX denoising was chosen \citep{Glasser2013, Jenkinson2002, Jenkinson2012, Fischl2012, Smith2013, Salimi2014, Griffanti2014}. For the comparison of the two approaches, $10$ sessions from $10$ different subjects were selected from the database. A Gaussian smoothing with a half width $FWHM = 5\ mm$ was then applied, using the \textit{SPM12} software package \footnote{\url{https://www.fil.ion.ucl.ac.uk/spm/software/spm12/}}, and the first five images were discarded to account for magnetic saturation effects.

%%%%%%%%%%%%%%%%%%%%%%%%%%%%%%%%
%%%%%%%%%%%%%%%%%%%%%%%%%%%%%%%%
\subsection{A hybrid cICA - EMD approach}

\begin{figure}[h!]
	\centering
    \includegraphics[width=120mm]{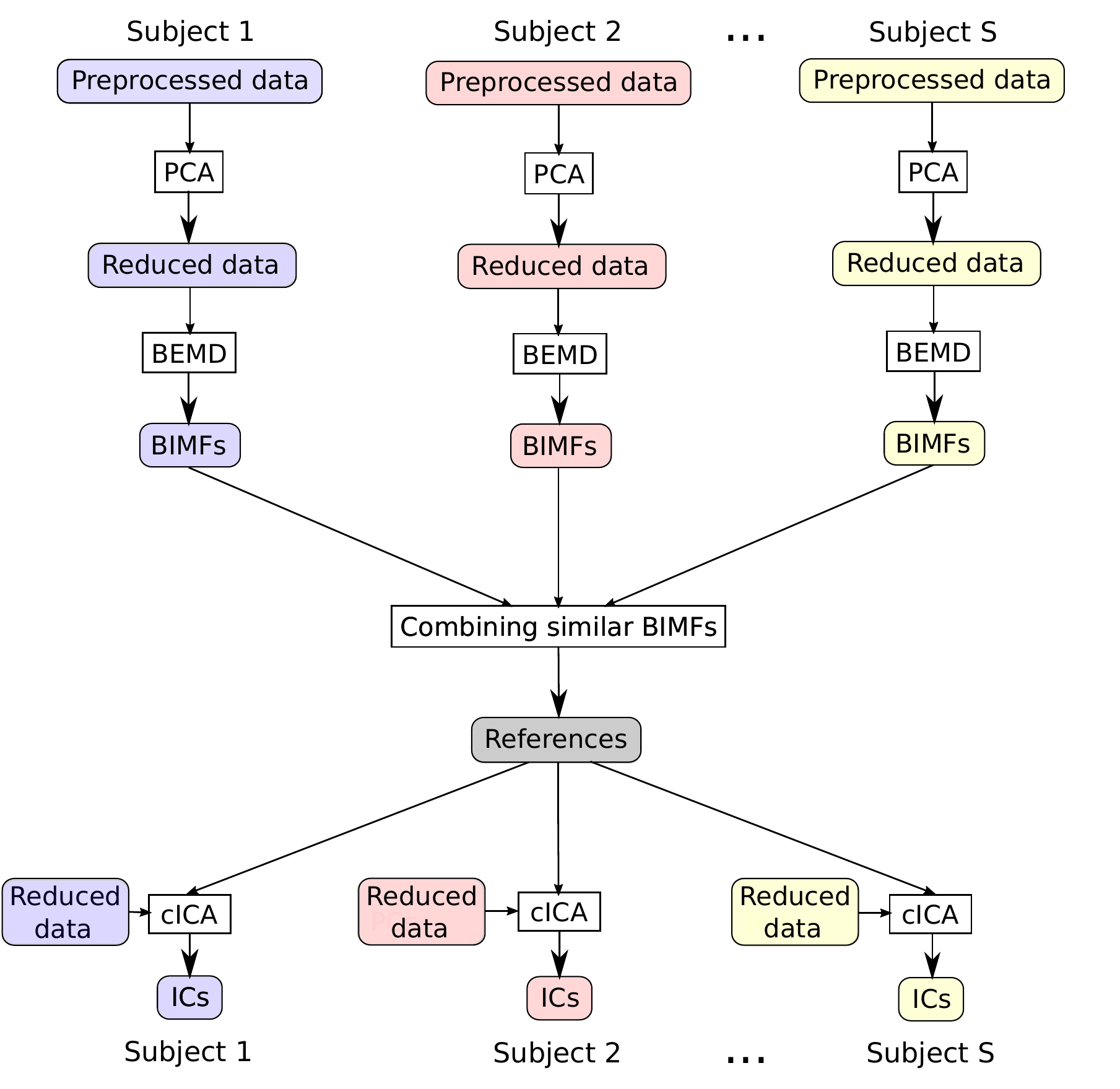}
    \caption{The flowchart sketches the main steps of the presented approach: First reducing the data with PCA, then extracting BIMFs with BEMD from the reduced data. Further these BIMFs of each subject are combined in order to get shared references for cICA, which finally help to obtain comparable ICs across subjects.}
    \label{fig:flowchart}
\end{figure}

In this section a new approach to deal with an ICA analysis across a group of subjects will be described. The flowchart in Figure \ref{fig:flowchart} presents an overview of the various steps of the data analysis. All processing steps were performed in  \textit{MATLAB 9.3 Release 2017b}.

%%%%%%%%%%%%%%%%%%%%%%%%%%%%%%%%
\subsubsection{Preprocessing}
The data as obtained from the data repository will be further pre-processed as explained in the following.
\begin{itemize}

\item
In a first  step, the voxel time series were linearly detrended and the voxel values were transformed to have zero mean and unit variance.
\item
Next a  mask common to all subjects was used to exclude voxels that were located outside of the brain. The mask was created, employing the \textit{GIFT} toolbox \footnote{\url{http://mialab.mrn.org/software/gift/}}, using the "Generate mask" option. 
\item
The data of subject $s$ is stored in a $K \times L$-dimensional matrix $\mb{X}^{(s)}$, containing in its columns $\mb{x}_l(k)$ the temporal evolution of $L$ brain voxels at $K$ time points.
\item
Then principal component analysis (PCA) related dimension reduction is performed, based on the singular value decomposition (SVD) of $\mb{X}^{(s)} = \mb{U}^{(s)} \bsb{\Sigma}^{(s)}  \mb{V}^{(s)T}$. If the row mean of $\mb{X}^{(s)}$ is removed, then $\textbf{U}^{(s)}$ contains the eigenvectors of the covariance matrix $\textbf{C}^{(s)} \propto \textbf{X}^{(s)}\textbf{X}^{(s)T} $ in its columns. The eigenvectors with largest eigenvalues indicate the directions of largest variance, which are denoted as principal components (PCs) \citep{Jolliffe2014}.
\item
Next, the fMRI images of each subject were projected onto the first $M = 20$ PCs $\mb{X}_M^{(s)} = (\mb{U}_M^{(s)})^T \, \mb{X}^{(s)}$. 
This reduces the number of images per session to $M = 20 < K$.
\item
Finally, from the reduced data sets the image slices were reconstructed to enter into the GiT-BEEMD analysis, while the reduced data $\mb{X}_M^{(s)}$ entered directly into the cICA processing path. The number of selected principal components $M$ determines the number of sources, estimated in the cICA step. A relatively low order of $M=20$ was chosen, to obtain robustly observed, large-scale resting state networks \citep{Heuvel2010}, and to make it easy to identify extracted networks, which are suitable for comparison of results in section \ref{sec_results}.
\eit

%%%%%%%%%%%%%%%%%%%%%%%%%%%%%%%%
\subsubsection{Green's function-based ensemble empirical mode decomposition}

For the next step, the extraction of suitable reference signals for cICA, the GiT-BEEMD technique was employed \citep{Albaddai2016}. The idea of this technique is to decompose a two-dimensional brain slice $I(\mb{r}) = I(x,y)$  into bi-dimensional intrinsic mode functions (BIMFs):

\beq
I(x,y) = \sum_{j=1}^{J} b_j(x,y)
\eeq
Here $b_j(x,y) $ denotes the $j$-th BIMF, which is estimated iteratively as described in Appendix \ref{bemd_algorithm}, and, in our notation, we include the residuum $r(x,y)$ as intrinsic mode $b_J(x,y)$. The first extracted BIMF contains the highest spatial frequency, which will decrease in every additionally extracted BIMF \citep{Albaddai2016}.

Each brain slice was decomposed into five intrinsic modes and one residuum by repeating the sifting step five times. The ensemble step was repeated only twice, whereby noise was either added or subtracted from the data once at each step. The assisting noise was generated with a noise amplitude of $a_{\eta} = 0.2 $. The tension parameter was initialized to $T_1 = 0.9$ and reduced after the extraction of the $j$-th BIMF $b_j$ to $T_{j+1} = T_j - \frac{1}{J}$. This avoids blob-like artifacts in low frequency modes, if the tension parameter is set too high \citep{Albaddai2016}. An example of a decomposition is provided in figure \ref{fig:example_BIMFs}. BIMFs with high spatial frequencies show highly localized spatial activation patterns which are spread all over the brain slice, while BIMFs with lower spatial frequencies concentrate the activity in specific areas in the brain. A combination of lowest spatial frequency intrinsic mode plus the residuum, i.~e. $b_5(x,y) + b_6(x,y) \equiv b_{56}(x,y)$, proved most appropriate to be used as a reference mode for cICA. From all decomposed two-dimensional slices, the corresponding modes $b_{56}(x,y)$ were organized into a three-dimensional data array, which then was concatenated into a 3D volume intrinsic mode function (VIMF). Decomposing the $M = 20$ brain volumes per subject in PC subspace results in $M$ VIMFs per subject. For the next processing steps the voxels inside of the brain are sorted into an $M \times L$ matrix again, denoted as $\mathbf{V}_M^{(s)}$. 

\begin{figure}[h!]
	\centering
    \includegraphics[width=120mm]{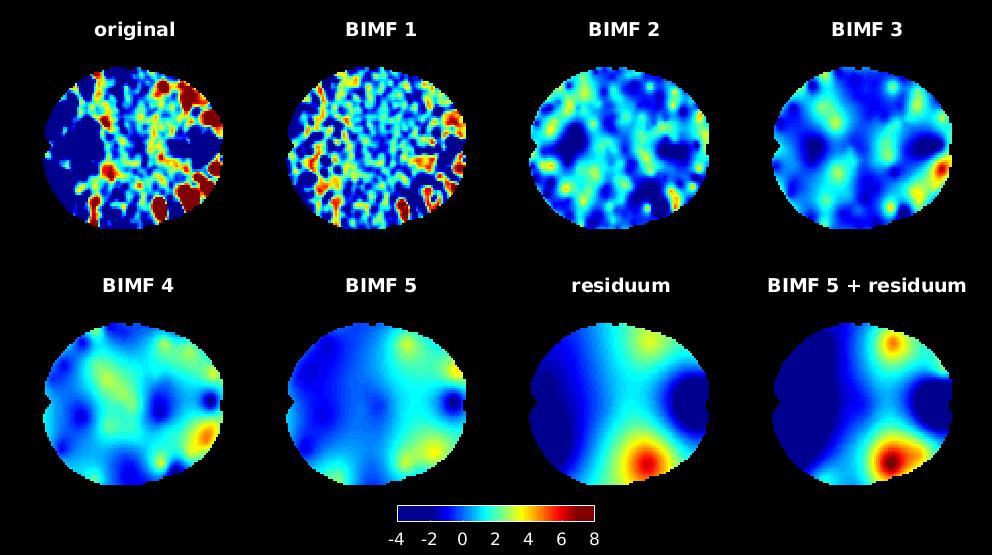}
    \caption{Example of a decomposition of a slice with GiT-BEEMD into 5 BIMFs and a residuum. Also the combination of fifth BIMF and residuum is illustrated. A mask was applied after the decomposition to set all intensity values that were located outside of the scanned brain to zero.}
    \label{fig:example_BIMFs}
\end{figure}
In order to extract from each subject RSNs, which are consistent across the proband cohort, corresponding intrinsic modes need to be identified. Averaging most similar modes between subjects then yields proper common reference signals for all subjects to be employed in a cICA of fMRI datasets. A visual inspection of the VIMFs of any two randomly chosen subjects showed that corresponding spatial patterns  might occur in different rows of $\mathbf{V}_M^{(s)}$. Hence first the extracted VIMFs need to be ordered according to their similarity between subjects. An efficient way to assign similar VIMFs between subjects is offered by the assignment algorithm proposed by \cite{Munkres1957}, based on the Hungarian method developed by \cite{Kuhn1955}. After computing the proper correspondences between the VIMFs of all subjects,  the references  $\mathbf{R}$, to be used in the cICA algorithm, then are obtained by averaging the corresponding VIMFs across all subjects. 

\clearpage

To summarize, the reference signals are computed as follows:

\begin{enumerate}
 \item  
 Initialize the reference as $\mb{R}=\mb{V}_M^{(1)}$
 \item  
 For $s=2, \ldots , S$, do:
    \begin{enumerate} 
    \item  
    Apply the Hungarian algorithm and re-order the rows of $\mb{V}_M^{(s)} \rightarrow \tilde{\mb{V}}_M^{(s)}  $
    \item  
    Update the reference $\mathbf{R}\leftarrow \frac{s-1}{s} \mb{R} +\frac{1}{s}\tilde{\mb{V}}_M^{(s)}$
    \end{enumerate}
\end{enumerate}
To apply the Hungarian algorithm, a cost function  $1-\rho(\mb{R}, \mb{V}_M^{(s)})$ is defined, which, if optimized, results in an ordering of the rows in $\mb{V}_M^{(s)}$ such that the sum of the correlation coefficients between pairs of rows in the two matrices is maximized. Note that the algorithm achieves the re-ordering without calculating all $M!$ possible assignments. Finally, each row of $\mb{R}$ is normalized, having entries with zero mean and unit variance, and $M = 20$ references for cICA algorithm are obtained. 
  
%%%%%%%%%%%%%%%%%%%%%%%%%%%%%%%%%%
\subsubsection{Constrained ICA}

Sources $\mb{Y}_M = [\mb{y}_1, \ldots , \mb{y}_L], \mb{y } \in \mbb{R}^M$ can be blindly estimated from the mixtures $\mb{X}_M = [\mb{x}_1, \ldots, \mb{x}_L], \mb{x} \in \mbb{R}^M$  according to:

\beq
\mb{Y}_M = \mb{W}\mb{X}_M
\eeq
with the demixing matrix defined as \(\mb{W} = [\mb{w}_1, \ldots,\mb{w}_M]^T$, where $\mb{w}^T_m$ are the rows of the demixing matrix and $\mb{X}_M$ collects all $L$ samples of the projected data after being spatially transformed to zero mean and unit variance.

Finding a demixing matrix is solved by designing an optimization problem where inequality and equality constraints are integrated in an augmented Lagrangian formulation. The inequality constraint terms in the Lagrange function are re-written as equality constraints with the help of a slack variable \citep{Lu2006}. After finding the optimal value of these slack variables, the modified version of the augmented Lagrangian function is written as

\begin{eqnarray}
 \mathcal{L}(\mathbf{W},\boldsymbol{\mu}) 
 %&=&   J(\mathbf{W}) + C(\mathbf{W}, \boldsymbol{\mu}) \nonumber \\
 &=& J(\mathbf{W}) + \sum_{m=1}^{M} \frac{1}{2\gamma_m} \left[ (\max\{0, \gamma_m h_m(\mathbf{w}^T_m)+ \mu_m\})^2 - \mu_m^2 \right]
\end{eqnarray} %maybe minus
where the $\mu_m$ are the Lagrangian parameters, while $\gamma_m$ represents a user defined penalty. The first term $ J(\mathbf{W}) $ reflects the cost function of ICA and the second term in the Lagrangian is related with the inequality constraint, which compares the $m$-th extracted component with the corresponding reference signal:

\beq
h(\mathbf{w}^T_m)=\varsigma_m - \epsilon(\mathbf{w}^T_m \mathbf{x},r_m)\leq 0
\eeq
where $\epsilon(\cdot)$ is a similarity measure and $\varsigma_m$ is a threshold parameter. Similarity is conventionally expressed either through a correlation measure $\mathbb{E}\{y_m r_m\}$ or the mean squared error $\mbb{E}\{(y_m - r_m)^2\}$, with $y_m = \mb{w}_m^T \mb{x}$. The expected value is approximated by an average over the available data. 
  
Estimating the demixing matrix $\mb{W}$, given the constraint introduced above, can be achieved in different ways, based either on negentropy-like costs functions (points 1 - 3) or on a maximum likelihood estimate (point 4):

\ben
    \item
    Simply one IC, most similar to the given reference signal, can be extracted. This approach can easily be extended to a multi-reference cICA. However, this additionally requires a decorrelation of the weights during each iteration to prevent different weights from converging to identical estimations \citep{Lu2006}.
    \item
    \cite{Lu2006} introduced an objective function for cICA,which contained an additional equality constraint to bound the weights. Later, a simplification was introduced by  \cite{Lin2007}, where equality constraints were omitted, rather the weight vectors were normalized at each iteration instead. 
    \item 
    Also cICA based on fixpoint learning \citep{Lin2009} was proposed, which should overcome the limitations of the second order Newton-like learning used in the cICA algorithm of \cite{Lu2006}.    
    \item 
    Finally, yet another version, using a cost function $J(\mb{W})$ based on a maximum likelihood estimate, has been proposed \citep{Rodriguez2014} according to
    
    \beq
	J(\mathbf{W}) \approx \mathbb{E} \left\{ \sum \limits_{m=1}^M \log ( p(\mathbf{w}_m^T \mathbf{x})) \right\}  + \log|\det(\mathbf{W})|
    \eeq
An iterative procedure is then derived to update the parameters $\boldsymbol{\mu}=[\mu_1, \ldots ,\mu_M]^T$ and the de-mixing matrix $\mathbf{W}$.  Thereby a decoupling scheme based on a Gram - Schmidt orthogonalization is proposed, finally yielding the following objective function

    \beq
    J(\mb{w}_m) \propto \mbb{E}\left\{ \log ( p(\mb{w}_m^T \mb{x}))\right \} + \log|\mb{d}_m^T\mathbf{w}_m|
    \eeq
where the decoupling vector $\mb{d}_m \in \mbb{R}^{M\times 1} $ is defined through  $\tilde{\mb{W}}_m \mb{d}_m=0$ and where $\tilde{\mb{W}}_m \in \mbb{R}^{(M-1)\times M}$ denotes the de-mixing matrix without entries to the $m$-th row. 
\een 

It has been shown  by \cite{Cardoso1997} that a maximum likelihood approach to ICA is equivalent to the seminal Infomax approach put forward by \cite{Bell1995}.  Thus, in analogy to the maximum likelihood approach, a constrained and decoupled version of the extended Infomax algorithm can be obtained \citep{Rodriguez2014}.  The extended Infomax algorithm is often used in an analysis of fMRI data \citep{Correa2007} and was also used in this study as the basis of the cICA. A more detailed description of this algorithm and a proper metacode are given in Appendix \ref{non-orth_cICA} for the convenience of the reader. 

The data, projected onto the first $M = 20$ PCs, and the $M$ references, transformed to zero mean and unit variance, together enter the cICA algorithm to finally extract $M = 20$ ICs. The weights are initialized with small random values, and the learning rate for the weights is set to $\eta = 0.5$. The scalar penalty can be set to 3 \citep{Rodriguez2014}. The influence of the references can be well determined by adjusting the threshold parameter. Therefore different settings have been studied, using the correlations $\mathbb{E}\{y_m r_m\}$ as distance measures, and results are presented in section \ref{sec_results}.

%%%%%%%%%%%%%%%%%%%%%%%%%%%%
%%%%%%%%%%%%%%%%%%%%%%%%%%%%

%%%%%%%%%%%%%%%%%%%%%%%%%%%%
%%%%%%%%%%%%%%%%%%%%%%%%%%%%
\subsection{Group-ICA}

Generally fMRI data is compared across a group of subjects by employing the gICA algorithm put forward by Calhoun and his group \citep{Calhoun2001}. This gICA is made available in the \textit{GIFT} toolbox \footnote{\url{http://mialab.mrn.org/software/gift/}} and was incorporated in the study for comparison. Voxel time series were preprocessed by variance normalization through linear detrending and transforming the data to zero mean and unit variance. The single subject data matrices $\textbf{X}^{(s)}_{K \times L}$ enter the first PCA step, with the temporal evolution of $L$ brain voxels at $K$ time points in columns. The subject datasets then were projected onto the first $ M' =1.5 \cdot M = 30 $ PCs in this step by applying an SVD to the data matrix.
Then the $S$ reduced $M' \times L$ matrices $\mb{X}_{M'}^{(s)}$ on subject level were concatenated to an \((S \cdot M') \times L \) group matrix $\textbf{X}_{S \cdot M'}^{(g)}$ entering a second PCA step. The group matrix is projected onto $ M = 20 $ PCs, resulting in a reduced $M \times L$ matrix $\mb{X}_M^{(g)}$. The spatial maps $\mb{S}_M$ are extracted from  $\mb{X}_M^{(g)} = \mb{A} \mb{S}_M$ by the extended Infomax algorithm \citep{Lee1999} and by additionally employing the ICASSO option \citep{Himberg2004}, running the ICA algorithm ten times with different initializations to assure greater stability. Subject specific spatial $\mb{S}_M^{(s)}$ maps were obtained by the GICA3 \citep{Erhardt2011} back-reconstruction approach. From these maps $M$ mean networks $\langle \mb{S}_{m*} \rangle, \; m=1, \ldots , M$ were obtained by averaging $\mb{S}_M^{(s)}$ over the subjects, i.~e. $\langle \mb{S}_M \rangle = \frac{1}{S}\sum_{s=1}^S \mb{S}_M^{(s)}$.

%%%%%%%%%%%%%%%%%%%%%%%%%%%
%%%%%%%%%%%%%%%%%%%%%%%%%%%
%%%%%%%%%%%%%%%%%%%%%%%%%%%
\section{Results} \label{sec_results}
The goal of the study was to compare RSNs obtained with the newly proposed cICA-EMD approach as opposed to RSNs resulting from the conventional gICA approach. RSNs denote functionally connected brain areas which, however, are anatomically separated but maintain a high level of activity in a resting state of the proband. They are represented in this study by the ICs extracted with the discussed techniques. In this study $M=20$ ICs were extracted with either method. Comparable RSNs, obtained by the different approaches, were identified by visual inspection and are depicted in figure \ref{fig:comparison}. There, references used for cICA are shown in the first row, while in the second row the ICs obtained therewith are presented, computed as mean ICs over subjects. In the third row of figure \ref{fig:comparison} the mean ICs obtained by gICA are exhibited. The significance of the resulting ICs was tested  with a one sample student's T-test by employing the \textit{SPM12} software package \footnote{\url{https://www.fil.ion.ucl.ac.uk/spm/software/spm12/}}. The resulting spatial maps of t-values are depicted in the fourth and fifth row in figure \ref{fig:comparison}. Spatial maps were thresholded at a significance level of $p<0.001$ $(t=4.30, df = 9)$. 

\begin{figure} [!htb]	
	\centering
	\includegraphics[width=1\textwidth]{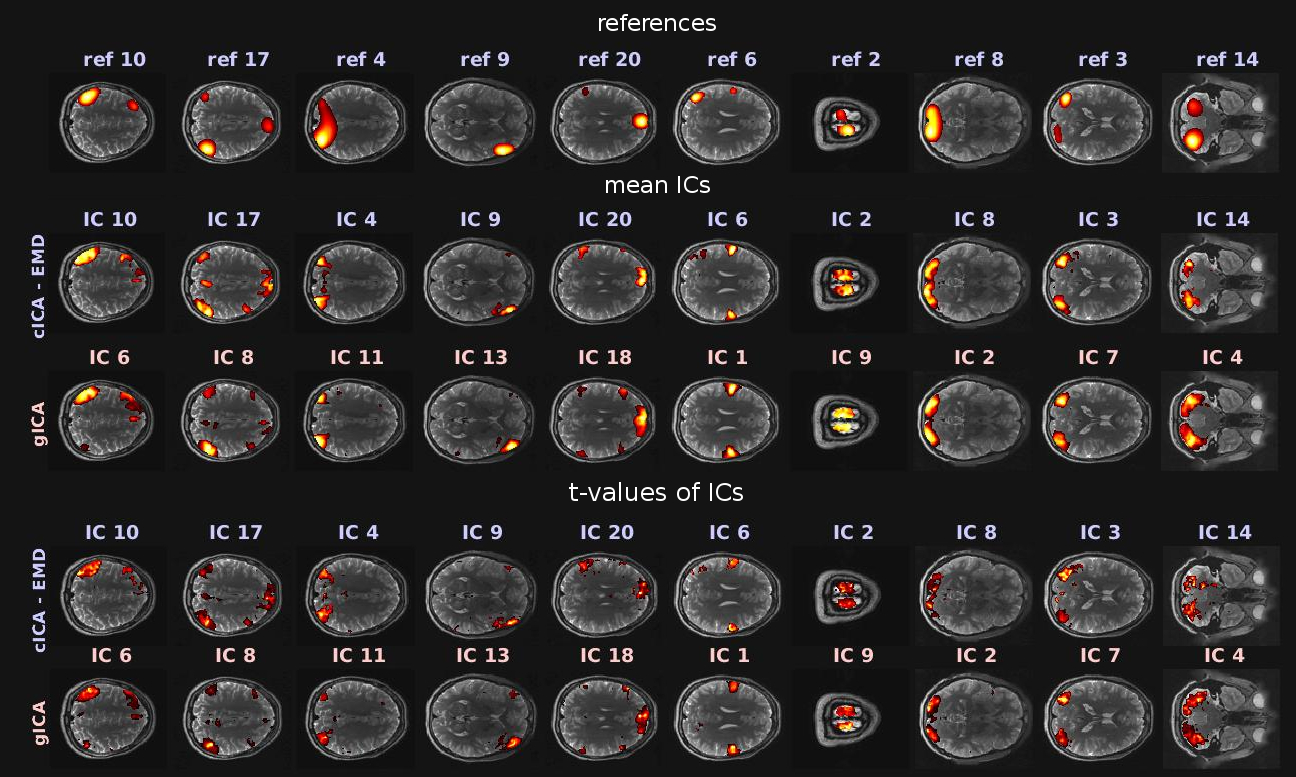}
	\caption[Comparison of results]{This figure illustrates RSNs, which were obtained by the two different approaches. The first row shows the references used for cICA, while the second row exhibits  mean ICs, averaged over the subject cohort and computed with the newly proposed cICA-EMD method. These ICs are contrasted in the third row with mean ICs obtained by gICA. In the fourth and fifth row of this figure, t-values of the RSNs can be found. All depicted slices were chosen such that they intersect the peak activation voxel of the corresponding ICs obtained by gICA. For visualization purposes, the activations of the networks shown in the first three rows are normalized to zero mean and unit variance. Furthermore, in accordance with common usage, the voxel intensities $\hat{I}(\mb{r})$ were thresholded by $\hat{I}(\mb{r}) > 2$, and in the sixth and ninth column the threshold was adjusted to $\hat{I}(\mb{r}) > 1.5$ for better recognizability of the networks. The color range of the heatmap was adjusted to the largest intensity value in every pictured slice. }
	\label{fig:comparison}		
\end{figure}

Most prominent brain areas are in IC 10/6 (obtained by cICA-EMD/gICA) the left inferior parietal lobule, in IC 17/8 the right angular/supramarginal gyrus, in IC 4/11 the superior occipital gyrus, in IC 9/13 the right inferior frontal gyrus, in IC 20/18 the anterior cingulate cortex, in IC 6/1 the precentral gyrus, in IC 2/9 the paracentral lobule, in IC 8/2 the middle occipital gyrus and in IC 3/7 the middle temporal gyrus. The obtained independent networks, representing well observed RSNs \citep{Heuvel2010}, and can be further grouped based on to their functions. The corresponding attentional and default mode networks are depicted in the first five columns, while the extracted auditory and sensorimotor networks are shown in the sixth and seventh column. Next, in the eighth and ninth column, visual networks are represented, and in the last column the cerebellum is shown. The similarity threshold for cICA was set to $\varsigma = 0.5$ in this example. All resting state networks obtained with both approaches, including the employed references are provided in the supplement.

The motivation of different group ICA approaches is to make this explorative analysis technique suitable for studies, where it is necessary to compare extracted networks between different subjects. This means issues with permutation indeterminacy and reproducibility of ICA have to be overcome, to obtain well comparable networks. Therefore a measure of interest for the evaluation of the two different approaches could be the consistency of activation patterns across subjects. This consistency was quantified by measuring how strong a resting state network $(\mb{y}^{(s)})_m^T \in \mbb{R}^L$ from one subject $s$ differs on average from the mean network $\langle \mb{y}\rangle^T_m = \frac{1}{S} \sum_{s=1}^{S} (\mb{y}^{(s)})_m^T $ across subjects. Pearson's correlation $\rho \big( (\mb{y}^{(s)})_m^T , \langle \mb{y} \rangle_m \big)$ was used to measure the correlation between standardized subject networks $\mb{y}_m^{(s)}$ and the related mean networks $\langle \mb{y} \rangle^T_m$. The following consistency measure is used: 

\beq \label{eq_con_per}
K_{per}(\mb{y}_m) = \frac{1}{S} \sum_{s=1}^{S} \rho \big( (\mb{y}^{(s)})_m^T , \langle \mb{y} \rangle_m \big) 
\eeq
This consistency measure was evaluated for all ICs obtained with the cICA-EMD approach for different settings of the similarity threshold $\varsigma$. An equivalent procedure was followed using the results from applying the gICA algorithm. The results are listed in table \ref{tab:consistency}. 

{\def\arraystretch{1.5}
\begin{table}[h!]
\caption[Consistency of ICs extracted by gICA]{Consistency values of ICs obtained by the proposed cICA-EMD approach for different settings of the similarity threshold $\varsigma$, as well as the values of ICs obtained by gICA. ICs were sorted like in figure \ref{fig:comparison}, so each column shows consistency values of comparable extracted networks.}
\vspace{5mm}
	\setlength{\tabcolsep}{5.0pt}
    cICA-EMD \\ 
    \begin{tabular}{l l l l l l l l l l l l}
    	\hline
        \textbf{threshold}\phantom{00} & \multicolumn{2}{l}{\textbf{IC \#}}\\
    	\cline{2-11}
     	& \textbf{10} & \textbf{17} & \textbf{4}\phantom{0} & \textbf{9}\phantom{0} & \textbf{20} & \textbf{6}\phantom{0} & \textbf{2}\phantom{0} & \textbf{8} & \textbf{3}\phantom{0} & \textbf{14}\phantom{0} \\
     	\hline
     	
     	\textbf{0.40} & 0.62 & 0.58 & 0.63 & 0.59 & 0.55 & 0.56 & 0.60 & 0.54 & 0.60 & 0.53\\
     	 
     	\textbf{0.50} & 0.67 & 0.68 & 0.68 & 0.68 & 0.66 & 0.66 & 0.69 & 0.65 & 0.70 & 0.59 \\

     	\textbf{0.60} & 0.71 & 0.76 & 0.73 & 0.74 & 0.71 & 0.72 & 0.71 & 0.73 & 0.73 & 0.62 \\

   \end{tabular} 
   \\
   \\
   \\
   gICA \\ 
    \begin{tabular}{l l l l l l l l l l l l}   
	    \hline
    	& \multicolumn{2}{l}{\textbf{IC \#}}\\
    	\cline{2-11}
    	\phantom{\textbf{threshold}}\phantom{00} & \textbf{6}\phantom{0} & \textbf{8}\phantom{0} & \textbf{11} &  \textbf{13} & \textbf{18} & \textbf{1}\phantom{0} & \textbf{9}\phantom{0} & \textbf{2}\phantom{0} & \textbf{7}\phantom{0} & \textbf{4}\phantom{0} \\
    	\cline{2-11}
    
     	& 0.73 & 0.65 & 0.68 & 0.70 & 0.66 & 0.66 & 0.66 & 0.70 & 0.71 & 0.60 \\
    
    \end{tabular}  
   \label{tab:consistency}    
\end{table}
} 
%new
By adjusting the threshold parameter $\varsigma$, it is possible to well determine the influence of the constraint during the optimization, so choosing a smaller threshold allows for more variability in the estimated components across subjects. Increasing the threshold increases the similarity between subject specific components and common references. This means that if the similarity between every subject component and the shared reference increases, also the similarity of components across subjects will increase, what is quantified by the consistency measure in equation \ref{eq_con_per}. Figure \ref{fig:consistency} illustrates the behavior of the consistency in dependence of the similarity threshold $\varsigma$ in comparison to gICA. If the threshold parameter is set to a value of $\varsigma = 0.40$, the consistency is lower in comparison to gICA. By further increasing  this threshold to a value of $\varsigma = 0.60$ the consistency of estimated resting state networks with the proposed approach exceeds that of gICA.

\begin{figure} [!htb]	
	\centering
	\includegraphics[width=0.9\textwidth]{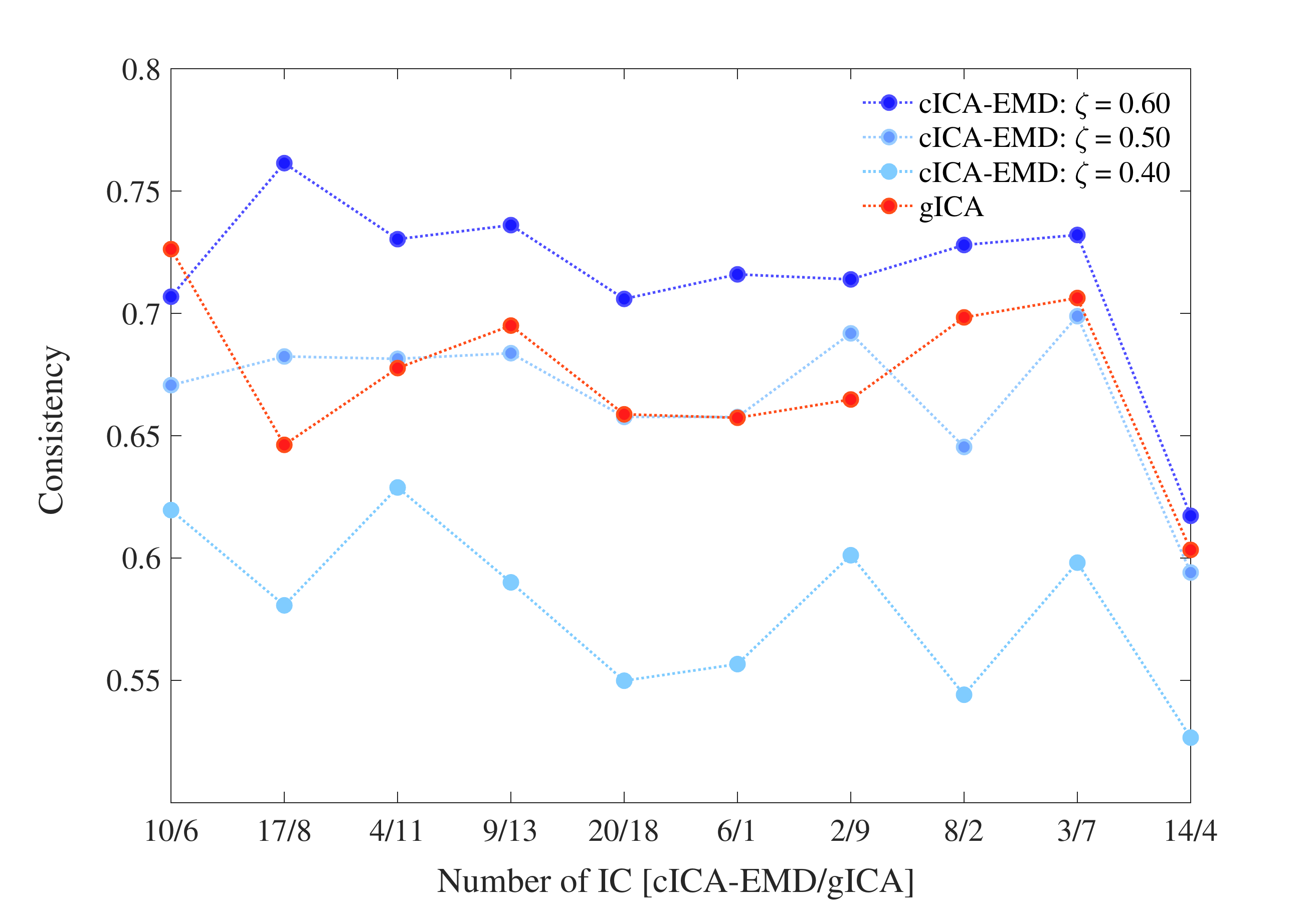}
	\caption[Consistency]{The figure illustrates the consistency values of the respective ICs. The numbers on the x-axis refer to the IC number of the cICA-EMD versus the gICA approach.}
	\label{fig:consistency}		
\end{figure}

%%%%%%%%%%%%%%%%%%%%%%%%%%%
%%%%%%%%%%%%%%%%%%%%%%%%%%%
%%%%%%%%%%%%%%%%%%%%%%%%%%%
\section{Discussion}

The motivation of this paper was to propose a novel workflow for extracting resting state networks consistent across a group of subjects. At first the dimensionality of the fMRI dataset was reduced at subject level with PCA. From that data intrinsic modes were extracted by employing the GiT-BEEMD algorithm \citep{Albaddai2016}. These subject-specific intrinsic modes reflect spatial activity patterns at different spatial frequencies. Hence, for each underlying spatial frequency a common reference mode can be formed. It turned out that low frequency modes concentrate the activity into spatially contiguous patterns and were especially well suited to serve as reference modes for the extraction of independent components with a cICA algorithm. Note that if intrinsic spatial modes, which are naturally ordered according to their dominant local spatial frequency, are chosen as reference signals within a cICA, the resulting independent modes are also ordered in correspondence to their assigned intrinsic modes. Thus, the natural ordering of the intrinsic modes with respect to their spatial frequencies helps to overcome the permutation ambiguity of ICA in extracting consistent resting state networks across subjects. It was demonstrated that our fMRI data processing pipeline produces commonly observed resting state patterns. These functional networks were then compared to those obtained by the widely used gICA, based on a temporal concatenation of individual datasets \citep{Calhoun2001_2}. It was shown that with the constrained extended Infomax algorithm \citep{Rodriguez2014}, the influence of the references upon the estimation of the related ICs could be well controlled. Based on the mathematically well described augmented Lagrangian framework in our workflow, it is transparent to the user with respect to how homologous resting state networks across subjects are deduced. 
%Even though the gICA, based on temporal concatenation, is a reliable tool for extracting resting state networks, it is not clear a priori, how the individual subject data sets will contribute to the estimation of common sources. 
In the presented processing pipeline, the cICA-EMD approach also allowed to shape the optimization procedure by adjusting the threshold parameter, which determines the impact of the reference onto the IC extraction. Choosing a lower threshold, e.g. allowing for a lower similarity to the references, more freedom could be given to the exploratory character of ICA and the formation of subject specific features. Defining a high threshold resulted, across subjects, in a higher consistency of the extracted resting state networks. These RSNs were even more consistent than those obtained by the conventional gICA. Although there is no exact ground truth on how resting state networks should ideally look like, the threshold can be chosen in a way such that the obtained networks optimally fulfill the requirements of a particular study. For example, when performing a classification task, the threshold can be chosen to maximize the accuracy of the classifier. So the good interpretability and high flexibility of the proposed processing pipeline can offer beneficial properties for applications in resting state studies.   

%%%%%%%%%%%%%%%%%%%%%%%%%%%
%%%%%%%%%%%%%%%%%%%%%%%%%%%
%%%%%%%%%%%%%%%%%%%%%%%%%%%
\section*{Conflict of Interest Statement}

The authors declare that the research was conducted in the absence of any commercial or financial relationship that could be construed as a potential conflict of interest.

%\section*{Author Contributions}
%
%S.W. and E.W.L. designed the study, S.W. and M.G. performed the simulations, S.W., E.W.L., A.M.T. and M.G wrote the manuscript, M.W.G. and E.W.L. supervised the study.

\section*{Acknowledgments}
Data were provided by the Human Connectome Project, WU-Minn Consortium (Principal Investigators: David Van Essen and Kamil Ugurbil; 1U54MH091657) funded by the 16 NIH Institutes and Centers that support the NIH Blueprint for Neuroscience Research; and by the McDonnell Center for Systems Neuroscience at Washington University. The authors also thank Saad Al-Baddai for helpful discussions.

%%%%%%%%%%%%%%%%%%%%%%%%%%%
%%%%%%%%%%%%%%%%%%%%%%%%%%%
%%%%%%%%%%%%%%%%%%%%%%%%%%%
\bibliography{cICA_EMD}

\begin{thebibliography}{45}
\providecommand{\natexlab}[1]{#1}
\providecommand{\url}[1]{\texttt{#1}}
\expandafter\ifx\csname urlstyle\endcsname\relax
  \providecommand{\doi}[1]{doi: #1}\else
  \providecommand{\doi}{doi: \begingroup \urlstyle{rm}\Url}\fi

\bibitem[Al-Baddai et~al.(2016{\natexlab{a}})Al-Baddai, Al-Subari, Tom{\'e},
  Ludwig, Salas-Gonzales, and Lang]{Albaddai2016_2}
S.~Al-Baddai, K.~Al-Subari, A.~M. Tom{\'e}, B.~Ludwig, D.~Salas-Gonzales, and
  E.~W. Lang.
\newblock {Analysis of fMRI images with bi-dimensional empirical mode
  decomposition based-on Green's functions}.
\newblock \emph{Biomedical Signal Processing and Control}, 30:\penalty0 53 --
  63, 2016{\natexlab{a}}.
\newblock ISSN 1746-8094.
\newblock \doi{10.1016/j.bspc.2016.06.019}.

\bibitem[Al-Baddai et~al.(2016{\natexlab{b}})Al-Baddai, Al-Subari, Tom{\'e},
  Sol-Casals, and Lang]{Albaddai2016}
S.~Al-Baddai, K.~Al-Subari, A.~M. Tom{\'e}, J.~Sol-Casals, and E.~W. Lang.
\newblock {A Green’s function-based Bi-dimensional empirical mode
  decomposition}.
\newblock \emph{Information Sciences}, 348:\penalty0 305 -- 321,
  2016{\natexlab{b}}.
\newblock ISSN 0020-0255.
\newblock \doi{10.1016/j.ins.2016.01.089}.

\bibitem[Allen et~al.(2011)Allen, Erhardt, Damaraju, Gruner, Segall, Silva,
  Havlicek, Rachakonda, Fries, Kalyanam, Michael, Caprihan, Turner, Eichele,
  Adelsheim, Bryan, Bustillo, Clark, {Feldstein Ewing, Sarah W.}, Filbey, Ford,
  Hutchison, Jung, Kiehl, Kodituwakku, Komesu, Mayer, Pearlson, Phillips,
  Sadek, Stevens, Teuscher, Thoma, and Calhoun]{Allen2011}
E.~A. Allen, E.~B. Erhardt, E.~Damaraju, W.~Gruner, J.~M. Segall, R.~F. Silva,
  M.~Havlicek, S.~Rachakonda, J.~Fries, R.~Kalyanam, A.~M. Michael,
  A.~Caprihan, J.~A. Turner, T.~Eichele, S.~Adelsheim, A.~D. Bryan,
  J.~Bustillo, V.~P. Clark, {Feldstein Ewing, Sarah W.}, F.~Filbey, C.~C. Ford,
  K.~Hutchison, R.~E. Jung, K.~A. Kiehl, P.~Kodituwakku, Y.~M. Komesu, A.~R.
  Mayer, G.~D. Pearlson, J.~P. Phillips, J.~R. Sadek, M.~Stevens, U.~Teuscher,
  R.~J. Thoma, and V.~D. Calhoun.
\newblock {A Baseline for the Multivariate Comparison of Resting-State
  Networks}.
\newblock \emph{Frontiers in Systems Neuroscience}, 5, 2011.
\newblock ISSN 16625137.
\newblock \doi{10.3389/fnsys.2011.00002}.

\bibitem[Allen et~al.(2014)Allen, Damaraju, Plis, Erhardt, Eichele, and
  Calhoun]{Allen2012}
E.~A. Allen, E.~Damaraju, S.~M. Plis, E.~B. Erhardt, T.~Eichele, and V.~D.
  Calhoun.
\newblock {Tracking whole-brain connectivity dynamics in the resting state.}
\newblock \emph{Cerebral cortex (New York, N.Y. : 1991)}, 24\penalty0
  (3):\penalty0 663--76, 2014.
\newblock ISSN 1460-2199.
\newblock \doi{10.1093/cercor/bhs352}.

\bibitem[Beckmann and Smith(2005)]{Beckmann2005}
C.~Beckmann and S.~Smith.
\newblock Tensorial extensions of independent component analysis for
  multisubject {fMRI} analysis.
\newblock \emph{NeuroImage}, 25\penalty0 (1):\penalty0 294 -- 311, 2005.
\newblock ISSN 1053-8119.
\newblock \doi{10.1016/j.neuroimage.2004.10.043}.

\bibitem[Bell and Sejnowski(1995)]{Bell1995}
A.~J. Bell and T.~J. Sejnowski.
\newblock An information-maximization approach to blind separation and blind
  deconvolution.
\newblock \emph{Neural Computation}, 7\penalty0 (6):\penalty0 1129--1159, Nov
  1995.
\newblock ISSN 0899-7667.
\newblock \doi{10.1162/neco.1995.7.6.1129}.

\bibitem[Biswal and Ulmer(1999)]{Biswal1999}
B.~B. Biswal and J.~L. Ulmer.
\newblock Blind source separation of multiple signal sources of {fMRI} data
  sets using independent component analysis.
\newblock \emph{Journal of computer assisted tomography}, 23 2:\penalty0
  265--71, 1999.

\bibitem[Calhoun et~al.(2001{\natexlab{a}})Calhoun, Adali, McGinty, Pekar,
  Watson, and Pearlson]{Calhoun2001}
V.~D. Calhoun, T.~Adali, V.~B. McGinty, J.~J. Pekar, T.~D. Watson, and G.~D.
  Pearlson.
\newblock {fMRI} activation in a visual-perception task: Network of areas
  detected using the general linear model and independent components analysis.
\newblock \emph{NeuroImage}, 14:\penalty0 1080--1088, 2001{\natexlab{a}}.

\bibitem[Calhoun et~al.(2001{\natexlab{b}})Calhoun, Adal{\i}, Pearlson, and
  Pekar]{Calhoun2001_2}
V.~D. Calhoun, T.~Adal{\i}, G.~D. Pearlson, and J.~J. Pekar.
\newblock A method for making group inferences from functional {MRI} data using
  independent component analysis.
\newblock \emph{Human brain mapping}, 14 3:\penalty0 140--51,
  2001{\natexlab{b}}.

\bibitem[Calhoun et~al.(2009)Calhoun, Liu, and Adal{\i}]{Calhoun2009}
V.~D. Calhoun, J.~Liu, and T.~Adal{\i}.
\newblock A review of group {ICA} for {fMRI} data and ica for joint inference
  of imaging, genetic, and {ERP} data.
\newblock \emph{NeuroImage}, 45 1 Suppl:\penalty0 S163--72, 2009.

\bibitem[Cardoso(1997)]{Cardoso1997}
J.~F. Cardoso.
\newblock Infomax and maximum likelihood for blind source separation.
\newblock \emph{IEEE Signal Processing Letters}, 4\penalty0 (4):\penalty0
  112--114, 1997.
\newblock \doi{10.1109/97.566704}.

\bibitem[Correa et~al.(2007)Correa, Adal{\i}, and Calhoun]{Correa2007}
N.~Correa, T.~Adal{\i}, and V.~D. Calhoun.
\newblock Performance of blind source separation algorithms for fmri analysis
  using a group ica method.
\newblock \emph{Magnetic Resonance Imaging}, 25\penalty0 (5):\penalty0 684 --
  694, 2007.
\newblock ISSN 0730-725X.
\newblock \doi{10.1016/j.mri.2006.10.017}.

\bibitem[Erhardt et~al.(2011)Erhardt, Rachakonda, Bedrick, Allen, Adal{\i}, and
  Calhoun]{Erhardt2011}
E.~B. Erhardt, S.~Rachakonda, E.~J. Bedrick, E.~A. Allen, T.~Adal{\i}, and
  V.~D. Calhoun.
\newblock {Comparison of multi-subject ICA methods for analysis of fMRI data.}
\newblock \emph{Human brain mapping}, 32 12:\penalty0 2075--95, 2011.

\bibitem[Esposito et~al.(2005)Esposito, Scarabino, Hyvarinen, Himberg,
  Formisano, Comani, Tedeschi, Goebel, Seifritz, and Salle]{Esposito2005}
F.~Esposito, T.~Scarabino, A.~Hyvarinen, J.~Himberg, E.~Formisano, S.~Comani,
  G.~Tedeschi, R.~Goebel, E.~Seifritz, and F.~D. Salle.
\newblock Independent component analysis of {fMRI} group studies by
  self-organizing clustering.
\newblock \emph{NeuroImage}, 25\penalty0 (1):\penalty0 193 -- 205, 2005.
\newblock ISSN 1053-8119.
\newblock \doi{10.1016/j.neuroimage.2004.10.042}.

\bibitem[Essen et~al.(2012)Essen, Ugurbil, Auerbach, Barch, Behrens, Bucholz,
  Chang, Chen, Corbetta, Curtiss, Penna, Feinberg, Glasser, Harel, Heath,
  Larson-Prior, Marcus, Michalareas, Moeller, Oostenveld, Petersen, Prior,
  Schlaggar, Smith, Snyder, Xu, and Yacoub]{Vanessen2012}
D.~V. Essen, K.~Ugurbil, E.~Auerbach, D.~Barch, T.~Behrens, R.~Bucholz,
  A.~Chang, L.~Chen, M.~Corbetta, S.~Curtiss, S.~D. Penna, D.~Feinberg,
  M.~Glasser, N.~Harel, A.~Heath, L.~Larson-Prior, D.~Marcus, G.~Michalareas,
  S.~Moeller, R.~Oostenveld, S.~Petersen, F.~Prior, B.~Schlaggar, S.~Smith,
  A.~Snyder, J.~Xu, and E.~Yacoub.
\newblock The human connectome project: A data acquisition perspective.
\newblock \emph{NeuroImage}, 62\penalty0 (4):\penalty0 2222 -- 2231, 2012.
\newblock ISSN 1053-8119.
\newblock \doi{10.1016/j.neuroimage.2012.02.018}.

\bibitem[Feinberg et~al.(2010)Feinberg, Moeller, M~Smith, Auerbach, Ramanna,
  Günther, F~Glasser, Miller, Ugurbil, and Yacoub]{Feinberg2010}
D.~Feinberg, S.~Moeller, S.~M~Smith, E.~Auerbach, S.~Ramanna, M.~Günther,
  M.~F~Glasser, K.~Miller, K.~Ugurbil, and E.~Yacoub.
\newblock Multiplexed echo planar imaging for sub-second whole brain fmri and
  fast diffusion imaging.
\newblock 5:\penalty0 e15710, 12 2010.

\bibitem[Fischl(2012)]{Fischl2012}
B.~Fischl.
\newblock Freesurfer.
\newblock \emph{NeuroImage}, 62\penalty0 (2):\penalty0 774 -- 781, 2012.
\newblock ISSN 1053-8119.
\newblock \doi{10.1016/j.neuroimage.2012.01.021}.
\newblock {20 YEARS OF fMRI}.

\bibitem[Fox and Raichle(2007)]{Fox2007}
M.~D. Fox and M.~E. Raichle.
\newblock {Spontaneous fluctuations in brain activity observed with functional
  magnetic resonance imaging}.
\newblock \emph{Nat Rev Neurosci}, 8\penalty0 (9):\penalty0 700--711, 2007.
\newblock ISSN 1471-003X.

\bibitem[Glasser et~al.(2013)Glasser, Sotiropoulos, Anthony~Wilson, Coalson,
  Fischl, L~Andersson, Xu, Jbabdi, Webster, Polimeni, Essen~DC, and
  Jenkinson]{Glasser2013}
M.~Glasser, S.~Sotiropoulos, J.~Anthony~Wilson, T.~Coalson, B.~Fischl,
  J.~L~Andersson, J.~Xu, S.~Jbabdi, M.~Webster, J.~Polimeni, V.~Essen~DC, and
  M.~Jenkinson.
\newblock The minimal preprocessing pipelines for the human connectome project.
\newblock 80:\penalty0 105, 10 2013.

\bibitem[Griffanti et~al.(2014)Griffanti, Salimi-Khorshidi, Beckmann, Auerbach,
  Douaud, Sexton, Zsoldos, Ebmeier, Filippini, Mackay, Moeller, Xu, Yacoub,
  Baselli, Ugurbil, Miller, and Smith]{Griffanti2014}
L.~Griffanti, G.~Salimi-Khorshidi, C.~F. Beckmann, E.~J. Auerbach, G.~Douaud,
  C.~E. Sexton, E.~Zsoldos, K.~P. Ebmeier, N.~Filippini, C.~E. Mackay,
  S.~Moeller, J.~Xu, E.~Yacoub, G.~Baselli, K.~Ugurbil, K.~L. Miller, and S.~M.
  Smith.
\newblock {ICA-based artefact removal and accelerated fMRI acquisition for
  improved resting state network imaging}.
\newblock \emph{NeuroImage}, 95:\penalty0 232 -- 247, 2014.
\newblock ISSN 1053-8119.
\newblock \doi{10.1016/j.neuroimage.2014.03.034}.

\bibitem[Harshman and Lundy(1994)]{Harshman1994}
R.~Harshman and M.~Lundy.
\newblock {PARAFAC}: parallel factor analysis.
\newblock \emph{Comput. Stat. Data Anal.}, 18:\penalty0 39 -- 72, 1994.

\bibitem[Himberg et~al.(2004)Himberg, Hyv{\"a}rinen, and Esposito]{Himberg2004}
J.~Himberg, A.~Hyv{\"a}rinen, and F.~Esposito.
\newblock Validating the independent components of neuroimaging time series via
  clustering and visualization.
\newblock \emph{NeuroImage}, 22 3:\penalty0 1214--22, 2004.

\bibitem[Hyv{\"a}rinen et~al.(2001)Hyv{\"a}rinen, Karhunen, and
  Oja]{Hyvarinen2001}
A.~Hyv{\"a}rinen, J.~Karhunen, and E.~Oja.
\newblock \emph{Independent Component Analysis}, volume~26.
\newblock 06 2001.
\newblock ISBN 9780471405405.

\bibitem[Jenkinson et~al.(2002)Jenkinson, Bannister, Brady, and
  Smith]{Jenkinson2002}
M.~Jenkinson, P.~Bannister, M.~Brady, and S.~Smith.
\newblock Improved optimization for the robust and accurate linear registration
  and motion correction of brain images.
\newblock \emph{NeuroImage}, 17\penalty0 (2):\penalty0 825 -- 841, 2002.
\newblock ISSN 1053-8119.
\newblock \doi{10.1006/nimg.2002.1132}.

\bibitem[Jenkinson et~al.(2012)Jenkinson, Beckmann, Behrens, Woolrich, and
  Smith]{Jenkinson2012}
M.~Jenkinson, C.~F. Beckmann, T.~E. Behrens, M.~W. Woolrich, and S.~M. Smith.
\newblock {FSL}.
\newblock \emph{NeuroImage}, 62\penalty0 (2):\penalty0 782 -- 790, 2012.
\newblock ISSN 1053-8119.
\newblock \doi{10.1016/j.neuroimage.2011.09.015}.
\newblock 20 YEARS OF fMRI.

\bibitem[Jolliffe(2014)]{Jolliffe2014}
I.~Jolliffe.
\newblock \emph{Principal Component Analysis}.
\newblock American Cancer Society, 2014.
\newblock ISBN 9781118445112.
\newblock \doi{10.1002/9781118445112.stat06472}.

\bibitem[Kuhn(1955)]{Kuhn1955}
H.~Kuhn.
\newblock The hungarian method for the assignment problem.
\newblock 2:\penalty0 83--98, 01 1955.

\bibitem[Lee et~al.(1999)Lee, Girolami, and Sejnowski]{Lee1999}
T.-W. Lee, M.~A. Girolami, and T.~J. Sejnowski.
\newblock Independent component analysis using an extended infomax algorithm
  for mixed sub-gaussian and super-gaussian sources.
\newblock \emph{Neural computation}, 11 2:\penalty0 417--41, 1999.

\bibitem[Lin et~al.(2007)Lin, Zheng, Yin, Liang, and Calhoun]{Lin2007}
Q.-H. Lin, Y.-R. Zheng, F.~Yin, H.~Liang, and V.~D. Calhoun.
\newblock A fast algorithm for one-unit {ICA-R}.
\newblock \emph{Inf. Sci.}, 177:\penalty0 1265--1275, 2007.

\bibitem[Lin et~al.(2009)Lin, Liu, Zheng, Liang, and Calhoun]{Lin2009}
Q.-H. Lin, J.~Liu, Y.-R. Zheng, H.~Liang, and V.~Calhoun.
\newblock Semiblind spatial {ICA of fMRI} using spatial constraints.
\newblock 31:\penalty0 1076--88, 07 2009.

\bibitem[Lu and Rajapakse(2006)]{Lu2006}
W.~Lu and J.~C. Rajapakse.
\newblock {ICA with Reference}.
\newblock \emph{Neurocomputing}, 69:\penalty0 2244--2257, 2006.

\bibitem[McKeown et~al.(1998)McKeown, Jung, Makeig, Brown, Kindermann, Lee, and
  Sejnowski]{McKeown1998}
M.~J. McKeown, T.~P. Jung, S.~Makeig, G.~E. Brown, S.~Kindermann, T.~W. Lee,
  and T.~J. Sejnowski.
\newblock Spatially independent activity patterns in functional {MRI} data
  during the stroop color-naming task.
\newblock \emph{Proceedings of the National Academy of Sciences of the United
  States of America}, 95 3:\penalty0 803--10, 1998.

\bibitem[Moeller et~al.(2010)Moeller, Yacoub, Olman, Auerbach, Strupp, Harel,
  and Ugurbil]{Moeller2010}
S.~Moeller, E.~Yacoub, C.~A. Olman, E.~Auerbach, J.~Strupp, N.~Y. Harel, and
  K.~Ugurbil.
\newblock Multiband multislice ge-epi at 7 tesla, with 16-fold acceleration
  using partial parallel imaging with application to high spatial and temporal
  whole-brain fmri.
\newblock \emph{Magnetic resonance in medicine}, 63 5:\penalty0 1144--53, 2010.

\bibitem[Munkres(1957)]{Munkres1957}
J.~Munkres.
\newblock Algorithms for the assignment and transportation problems.
\newblock \emph{Journal of the Society for Industrial and Applied Mathematics},
  5:\penalty0 32--38, 03 1957.

\bibitem[Nunes et~al.(2003)Nunes, Bouaoune, Del{\'e}chelle, Niang, and
  Bunel]{Nunes2003}
J.~C. Nunes, Y.~Bouaoune, {\'E}.~Del{\'e}chelle, O.~Niang, and P.~Bunel.
\newblock Image analysis by bidimensional empirical mode decomposition.
\newblock \emph{Image Vision Comput.}, 21:\penalty0 1019--1026, 2003.

\bibitem[Remes et~al.(2011)Remes, Starck, Nikkinen, Ollila, Beckmann, Tervonen,
  Kiviniemi, and Silven]{Remes2011}
J.~J. Remes, T.~Starck, J.~Nikkinen, E.~Ollila, C.~F. Beckmann, O.~Tervonen,
  V.~Kiviniemi, and O.~Silven.
\newblock {Effects of repeatability measures on results of fMRI sICA: a study
  on simulated and real resting-state effects.}
\newblock \emph{NeuroImage}, 56\penalty0 (2):\penalty0 554--69, 2011.
\newblock ISSN 1095-9572.
\newblock \doi{10.1016/j.neuroimage.2010.04.268}.

\bibitem[Rodriguez et~al.(2014)Rodriguez, Anderson, Li, and
  Adal{\i}]{Rodriguez2014}
P.~A. Rodriguez, M.~Z. Anderson, X.-L. Li, and T.~Adal{\i}.
\newblock General non-orthogonal constrained {ICA}.
\newblock \emph{IEEE Transactions on Signal Processing}, 62:\penalty0
  2778--2786, 2014.

\bibitem[Salimi-Khorshidi et~al.(2014)Salimi-Khorshidi, Douaud, Beckmann,
  Glasser, Griffanti, and Smith]{Salimi2014}
G.~Salimi-Khorshidi, G.~Douaud, C.~F. Beckmann, M.~F. Glasser, L.~Griffanti,
  and S.~M. Smith.
\newblock Automatic denoising of functional {MRI} data: Combining independent
  component analysis and hierarchical fusion of classifiers.
\newblock \emph{NeuroImage}, 90:\penalty0 449 -- 468, 2014.
\newblock ISSN 1053-8119.
\newblock \doi{10.1016/j.neuroimage.2013.11.046}.

\bibitem[Schmithorst and Holland(2004)]{Schmithorst2004}
V.~J. Schmithorst and S.~K. Holland.
\newblock Comparison of three methods for generating group statistical
  inferences from independent component analysis of functional magnetic
  resonance imaging data.
\newblock \emph{Journal of Magnetic Resonance Imaging}, 19\penalty0
  (3):\penalty0 365--368, 2004.
\newblock \doi{10.1002/jmri.20009}.

\bibitem[Setsompop et~al.(2012)Setsompop, Gagoski, Polimeni, Witzel, Wedeen,
  and Wald]{Setsompop2012}
K.~Setsompop, B.~Gagoski, J.~R. Polimeni, T.~Witzel, V.~J. Wedeen, and L.~L.
  Wald.
\newblock Blipped-controlled aliasing in parallel imaging for simultaneous
  multislice echo planar imaging with reduced g-factor penalty.
\newblock \emph{Magnetic resonance in medicine}, 67 5:\penalty0 1210--24, 2012.

\bibitem[Smith et~al.(2013)Smith, Beckmann, Andersson, Auerbach, Bijsterbosch,
  Douaud, Duff, Feinberg, Griffanti, Harms, Kelly, Laumann, Miller, Moeller,
  Petersen, Power, Salimi-Khorshidi, Snyder, Vu, Woolrich, Xu, Yacoub,
  Uğurbil, Essen, and Glasser]{Smith2013}
S.~M. Smith, C.~F. Beckmann, J.~Andersson, E.~J. Auerbach, J.~Bijsterbosch,
  G.~Douaud, E.~Duff, D.~A. Feinberg, L.~Griffanti, M.~P. Harms, M.~Kelly,
  T.~Laumann, K.~L. Miller, S.~Moeller, S.~Petersen, J.~Power,
  G.~Salimi-Khorshidi, A.~Z. Snyder, A.~T. Vu, M.~W. Woolrich, J.~Xu,
  E.~Yacoub, K.~Uğurbil, D.~C.~V. Essen, and M.~F. Glasser.
\newblock Resting-state {fMRI} in the human connectome project.
\newblock \emph{NeuroImage}, 80:\penalty0 144 -- 168, 2013.
\newblock ISSN 1053-8119.
\newblock \doi{10.1016/j.neuroimage.2013.05.039}.

\bibitem[Svens{\'e}n et~al.(2002)Svens{\'e}n, Kruggel, and Benali]{Svensn2002}
M.~Svens{\'e}n, F.~Kruggel, and H.~Benali.
\newblock {ICA of fMRI group study data}.
\newblock \emph{NeuroImage}, 16 3 Pt 1:\penalty0 551--63, 2002.

\bibitem[van~den Heuvel and Pol(2010)]{Heuvel2010}
M.~M. H.~P. van~den Heuvel and H.~E.~H. Pol.
\newblock Exploring the brain network: a review on resting-state fmri
  functional connectivity.
\newblock \emph{European neuropsychopharmacology : the journal of the European
  College of Neuropsychopharmacology}, 20 8:\penalty0 519--34, 2010.

\bibitem[Wessel and Bercovici(1998)]{Wessel1998}
P.~Wessel and D.~Bercovici.
\newblock Interpolation with splines in tension: A {G}reen's function approach.
\newblock 30:\penalty0 77--93, 01 1998.

\bibitem[Xu et~al.(2012)Xu, Moeller, Strupp, Auerbach, Chen, A.~Feinberg,
  Ugurbil, and Yacoub]{Xu2012}
J.~Xu, S.~Moeller, J.~Strupp, E.~Auerbach, L.~Chen, D.~A.~Feinberg, K.~Ugurbil,
  and E.~Yacoub.
\newblock Highly accelerated whole brain imaging using
  aligned-blipped-controlled-aliasing multiband {EPI}.
\newblock \emph{Proceedings of the 20th Annual Meeting of ISMRM}, page 2036,
  2012.

\end{thebibliography}

\clearpage

%%%%%%%%%%%%%%%%%%%%%%%%%%%
%%%%%%%%%%%%%%%%%%%%%%%%%%%
%%%%%%%%%%%%%%%%%%%%%%%%%%%
\begin{appendix}

\section{Appendix}

\subsection{A Green's function-based bi-dimensional empirical mode decomposition} \label{bemd_algorithm}

The process of extracting BIMFs $b_j(x,y)$  with a bi-dimensional EMD (BEMD) algorithm can be summarized in the following steps \citep{Albaddai2016_2, Nunes2003}:

\noindent\rule{\textwidth}{1pt}

\textit{Bi-dimensional empirical mode decomposition}

\noindent\rule[6pt]{\textwidth}{1pt}

\begin{enumerate} \itemsep0.5em

\item Choose the number of intrinsic modes $J$ and the number of sifting steps $N$ and set $r(x,y) = f(x,y)$

\item Extract the $j$-th BIMF by repeating the sifting steps $N$ times:

\begin{enumerate} \itemsep2mm

\item Identify all local maxima and minima of the array $r(x,y)$
\item Interpolate these local maxima to an upper envelope surface $s_{max}(x,y)$ and local minima to a lower envelope surface $ s_{min}(x,y) $ and calculate the mean between upper envelope surface and lower envelope surface $ s_{mean}(x,y)  = 0.5 \, ( s_{max}(x,y) + s_{min}(x,y) ) $
\item Update $r(x,y)$ with $r(x,y) \leftarrow r(x,y) - s_{mean}(x,y)$
\item If loop is finished, set $b_j(x,y) = r(x,y) $, otherwise repeat steps (a) - (d)
\end{enumerate}

\item Subtract all calculated BIMFs $b_{1...j}$ from $f(x,y)$ to obtain new $r(x,y) = f(x,y) - \sum_{j < j + 1} b_j(x,y)$

\item If all $J$ BIMFs are extracted, $r(x,y)$ is the residuum, otherwise repeat step 2 to compute the next BIMF $b_{j+1}$

\end{enumerate}

\noindent\rule[9pt]{\textwidth}{1pt}

Interpolation schemes, which are used to describe the upper and lower envelope surface, usually suffer from problems like \textit{computational load}, \textit{boundary artefacts} and \textit{over- and undershooting} \citep{Albaddai2016}. Using a Green's function-based interpolation scheme,  local maxima or minima can be considered as the known points of the envelope surface, which can be found with an 8-connected neighborhood strategy. Then surface envelopes $s(\textbf{r}_{u})$, at Cartesian coordinates $\textbf{r}_{u} = [x_u, y_u]^T$, are represented as a weighted sum of Green's functions \citep{Wessel1998}:

\beq \label{surface}
s(\textbf{r}_{u}) = \sum_{n=1}^{N} v_{n} \Phi(\textbf{r}_u,\textbf{r}_n ) ,
\eeq
where $\Phi(\textbf{r}_u,\textbf{r}_n)$ represent the Green's functions and $v_{n}$ the corresponding weights. Further $\textbf{r}_u$ denotes a point where the surface is unknown and $\textbf{r}_n$ describes the $n$-th constraint, which corresponds to a local extremum. The Green's function, expressed in 2D-Cartesian coordinates, reads as:

\beq
\Phi(\textbf{r}_u,\textbf{r}_n) = \mbox{log}(p|\textbf{r}_u - \textbf{r}_n|) + K_0(p|\textbf{r}_u - \textbf{r}_n|) 
\eeq
with $K_0( \cdot )$ representing the modified Bessel function of the second kind and zero order and $|\cdot|$ the Euclidean distance. Here $p^2 = \frac{T}{D}$ is related with tension $T$ at the boundaries, and \(D\) describes the flexural rigidity of the surface \citep{Wessel1998}. The estimation of the envelope surface is based on two steps. In a first step weights $v_n$ can be estimated by taking the known values of local maxima (or minima) as the values $s(\textbf{r}_{n}) = [s(\textbf{r}_1), \,.\,.\,.\, ,s(\textbf{r}_N)]^T$ in a total of $N$ locations $ \textbf{r}_{n}$ and solving a linear system of $N$ equations, described by equation \ref{surface}. In a second step, if the weights $v_n$ are now obtained, the surface can be estimated at any point $\textbf{r}_u $. 

To avoid mode mixing and boundary artifacts, a noise assisted ensemble version of the Green's function based BEMD (GiT-BEEMD) can be used \citep{Albaddai2016_2}. Adding and subtracting noise $\eta$ from the original image $f(x,y)$ leads to two noisy versions $\tilde{f}(x,y)^*: \tilde{f}^+(x,y) = f(x,y) + \eta$ and $\tilde{f}^-(x,y) = f(x,y) - \eta$.  Both versions $\tilde{f}(x,y)^*$ can now be  decomposed into BIMFs. By computing the mean as $0.5\,(\tilde{f}^+(x,y) + \tilde{f}^-(x,y) )$ the original array $f(x,y)$ could be reconstructed and therefore after decomposing the two versions $ \tilde{f}(x,y)^* $, the BIMFs of $f(x,y)$ can be naturally obtained by averaging BIMFs of the noisy versions. For this version of BEMD it is sufficient to use a few ensemble steps only to improve the image quality significantly \citep{Albaddai2016}, reducing computational load.  The two-dimensional image decomposition was applied to the slices of volumetric fMRI images in the transverse anatomical plane.

%%%%%%%%%%%%%%%%%%%%%%%%%%%%%%%%
%%%%%%%%%%%%%%%%%%%%%%%%%%%%%%%%
\subsection{Non-orthogonal constrained extended Infomax} \label{non-orth_cICA}

The objective function for ICA, based on Maximum Likelihood, can be derived as \citep{Hyvarinen2001}
\beq
J(\mathbf{W}) \approx \mathbb{E} \Big\{ \sum \limits_{m=1}^M \log ( p(\mathbf{w}_m^T \mathbf{x})) \Big\} + \log|\det(\mathbf{W})|
\label{eqML}
\eeq
It is proposed a decoupling strategy for the second term of cost function resulting in a objective function for each of the rows of the mixing matrix \citep{Rodriguez2014}
\beq
J(\mathbf{w}_m) \approx \mathbb{E}\left\{ \log ( p(\mathbf{w}_m^T \mathbf{x}))\right \} + \log|(\mathbf{d}_m^T\mathbf{w}_m)| + \mbox{log} \,(S)
\label{eqML}
\eeq
with $ S = \sqrt{| \mbox{det} \left( \tilde{\textbf{W}}_m \tilde{\textbf{W}}_m^T \right) |} $, where $\tilde{\mathbf{W}}_m$ is the de-mixing matrix without $m-th$ row. The decoupling vector is the vector $\tilde{\mathbf{W}}_m\mathbf{d}_m=0$ and can be computed as
\beq
\mathbf{d}_m=(\mathbf{I}-\tilde{\mathbf{W}}_m^T(\tilde{\mathbf{W}}_m\tilde{\mathbf{W}}^T_m)^{-1} \tilde{\mathbf{W}}_m)\mathbf{v}
\eeq
where $\mathbf{v}$ is a vector with Gaussian random values. The vector gradient of this function is then derived as
\beq
\nabla_{\mathbf{w}_m} J(\mathbf{w}_m)=\mathbb{E}\left\{f_m(\mathbf{w}_m^T \mathbf{x})\mathbf{x}^T\right \}+\frac{\mathbf{d}_m^T}{\mathbf{d}_m^T\mathbf{w}_m}
\eeq
with so-called score functions defined as:

\beq
f_m (y_m) = \frac{\partial \, \mbox{log} \, p(y_m)}{\partial y_m} 
\eeq
The extended Infomax algorithm defines these non-linearities $ f_m (y_m) $ differently for sub-Gaussian or super-Gaussian components \citep{Lee1999}. The Lagrangian multipliers $ \mu_m $ are updated by gradient ascent

\beq
\mu_m  \leftarrow \mbox{max}  \{0, \gamma_m h_m(y_m,r_m) + \mu_m \}
\eeq
For the algorithm a convergence metric like $ [ (vec(\Delta \textbf{W}))^T vec(\Delta \textbf{W})] <  \tau $ can be adapted \citep{Rodriguez2014}. Here $ \Delta \textbf{W} $ is the element by element difference of $ \textbf{W} $ after each iteration, $ vec(\cdot) $ stores all elements of a matrix in a column vector and $ \tau $ is the tolerance value.
So the non-orthogonal constrained extended Infomax can be summarized in the following steps:

\noindent\rule{\textwidth}{1pt}

\textit{Non-orthogonal constrained extended Infomax}

\noindent\rule[6pt]{\textwidth}{1pt}

\begin{enumerate} \itemsep0.5em 

\item
Randomly initialize \textbf{W} and initialize $\mu_m$, set $\gamma_m$ and thresholds $\varsigma_m$

\item 
\textbf{for} weights $\textbf{w}_m, \; m = 1, \ldots, M$ do:

\begin{enumerate} \itemsep2.5mm
% \vspace{-1pt}
\item 
Compute the vector $\mb{d}_m=(\mb{I}-\tilde{\mb{W}}_m^T(\tilde{\mb{W}}_m\tilde{\mb{W}}_m^T)^{-1} \tilde{\mb{W}}_m)\mb{v}$, with $\textbf{d}_m \perp \textbf{w}_{i \neq m}$.  Here $\mb{v}$ is a Gaussian random vector.

\item
Compute $\mb{y}_m = \mb{w}_m^T \mb{x}$

\item 
Update $ \mu_m  \leftarrow \mbox{max} \{0, \gamma_m h_m(y_m,r_m) + \mu_m \} $

\item 
Let  $ \Delta \mb{w}^T_m \propto \frac{\mb{d}_m^T}{\mb{d}_m^T \mb{w}_m} + \mbb{E} \{f_m(\mb{w}_m^T \mb{x})\mb{x}^T\} - \frac{1}{2} \, \mu_m \mbb{E}\{ h_m'(y_m,r_m) \mb{x}^T\} $ 

and set $f_m(y_m) = \tanh (y_m) - y_m $ for sub-Gaussian sources and 

$f_m(y_m) = - \tanh(y_m) - y_m $ for super-Gaussian sources.

\item 
Update $ \mb{w}_m \leftarrow \mb{w}_m + \Delta \mb{w}_m $

\item 
And normalize $ \mb{w}_m \leftarrow \frac{\mb{w}_m}{\parallel \mb{w}_m \parallel}$

\end{enumerate}
\textbf{end for}

\item 
Repeat step 2 until convergence.

\end{enumerate}

\noindent\rule[9pt]{\textwidth}{1pt}

\end{appendix}

\clearpage

%%%%%%%%%%%%%%%%%%%%%%%%%%%
%%%%%%%%%%%%%%%%%%%%%%%%%%%
%%%%%%%%%%%%%%%%%%%%%%%%%%%
\section{Supplementary material}

\subsection{References for cICA}

\begin{figure}[h!]
	\centering
	\begin{subfigure}[b]{0.495\textwidth} 
    	\includegraphics[width=\textwidth]{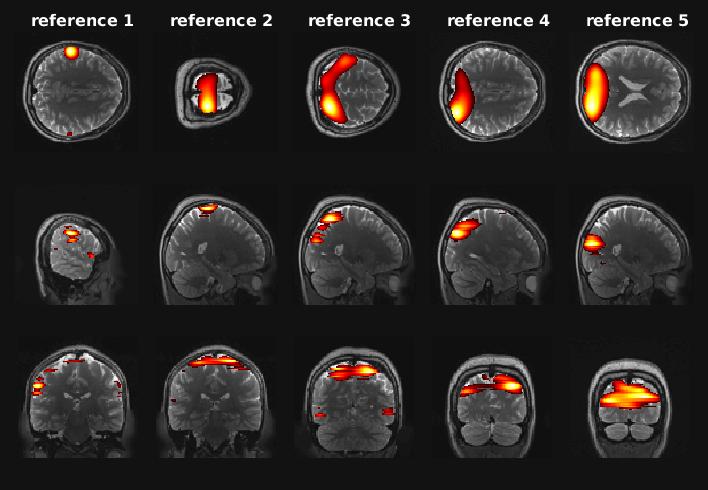}
	\end{subfigure}\hspace{-0.1em}%
	\begin{subfigure}[b]{0.495\textwidth} 
    	\includegraphics[width=\textwidth]{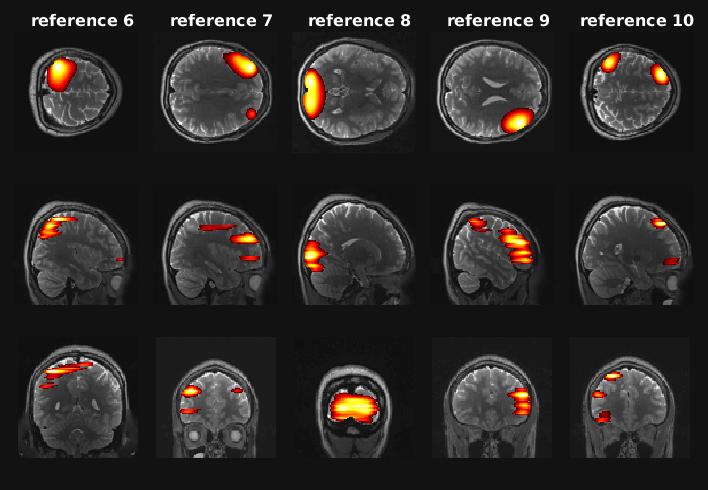}
	\end{subfigure} 

	\begin{subfigure}[b]{0.495\textwidth} 
    	\includegraphics[width=\textwidth]{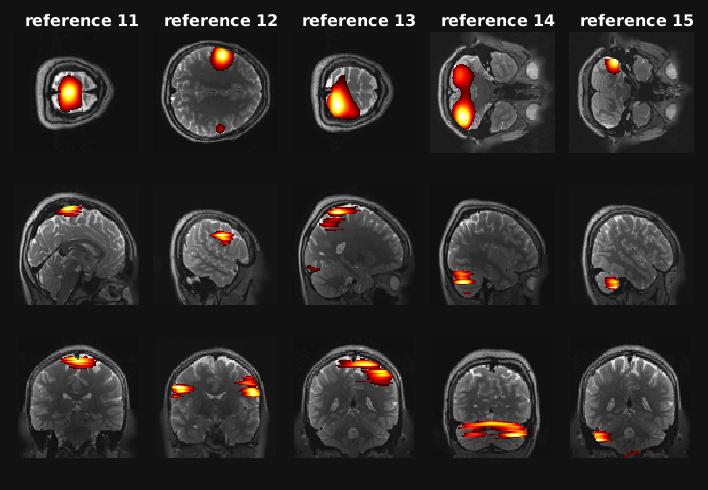}
	\end{subfigure}\hspace{-0.1em}%
	\begin{subfigure}[b]{0.495\textwidth} 
    	\includegraphics[width=\textwidth]{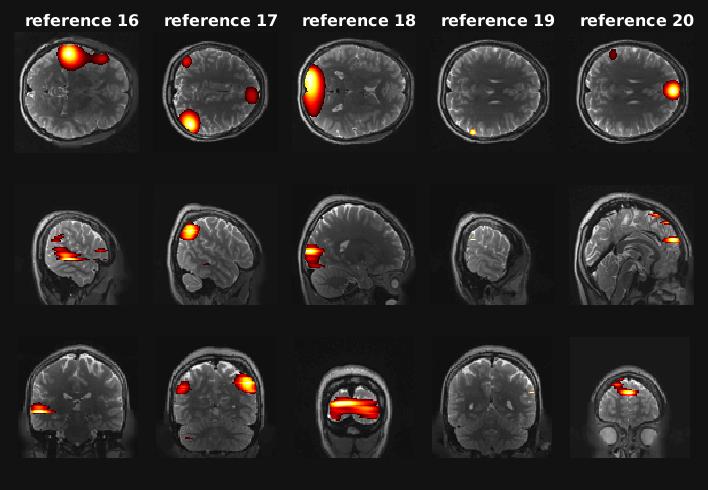}
	\end{subfigure}
	\caption{References used for cICA. The references are presented in the three principal anatomical planes: transverse, sagittal and frontal. The three planes intersect the voxel with highest intensity value in every volumetric image. For visualization purposes, the activations of the networks shown are normalized to zero mean and unit variance, and the intensity of the pictured brain slice $\hat{I}(\mb{r})$ was thresholded by $\hat{I}(\mb{r}) > 2$. The color range is adjusted to the largest intensity value in every pictured slice.}
\end{figure}

\clearpage

\subsection{ICs obtained by the cICA - EMD approach}

\begin{figure}[h!]
	\centering
	\begin{subfigure}[b]{0.495\textwidth} 
    	\includegraphics[width=\textwidth]{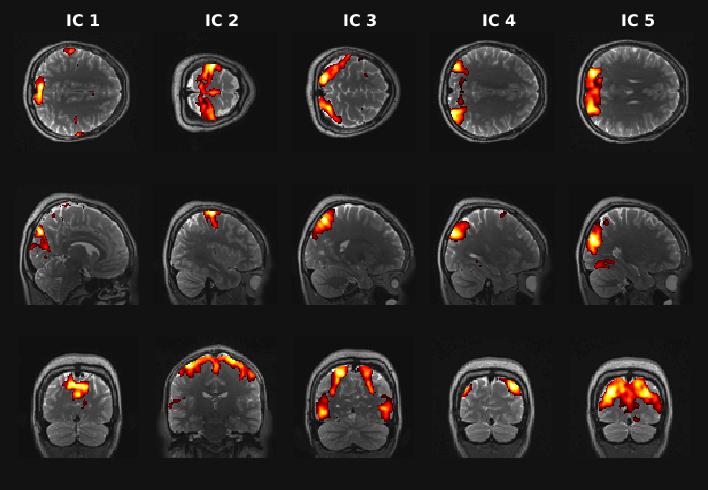}
	\end{subfigure}\hspace{-0.1em}%
	\begin{subfigure}[b]{0.495\textwidth} 
    	\includegraphics[width=\textwidth]{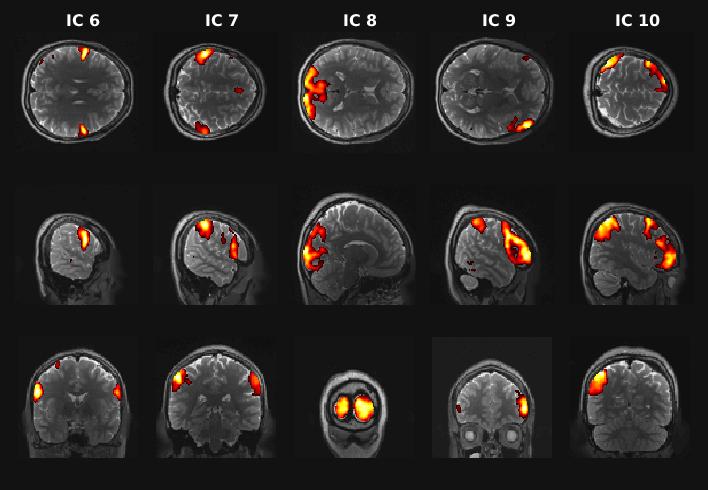}
	\end{subfigure} 

	\begin{subfigure}[b]{0.495\textwidth} 
    	\includegraphics[width=\textwidth]{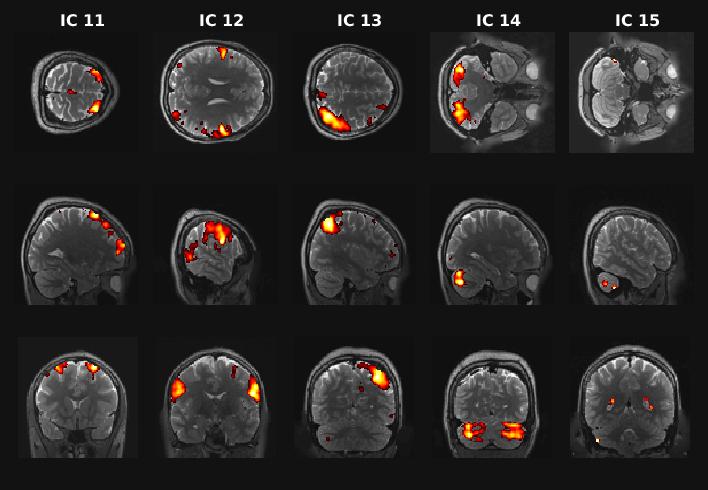}
	\end{subfigure}\hspace{-0.1em}%
	\begin{subfigure}[b]{0.495\textwidth} 
    	\includegraphics[width=\textwidth]{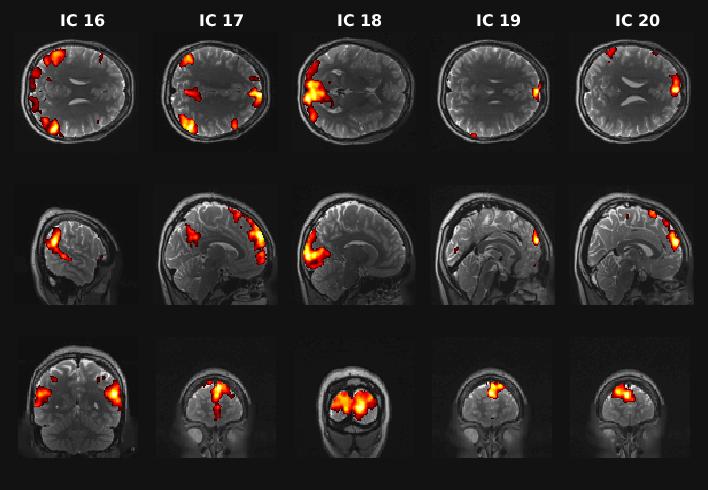}
	\end{subfigure}
	\caption{ICs obtained by the proposed approach. The depicted ICs are computed as mean ICs over subjects. Here the similarity threshold for cICA was set to $\mbb{\varsigma}_m = 0.5$. The ICs are presented in the three principal anatomical planes: transverse, sagittal and frontal. The three planes intersect the voxel with highest intensity value in every volumetric image. For visualization purposes, the activations of the networks shown are normalized to zero mean and unit variance, and the intensity of the pictured brain slice $\hat{I}(\mb{r})$ was thresholded by $\hat{I}(\mb{r}) > 2$. The color range is adjusted to the largest intensity value in every pictured slice.}
\end{figure}

\clearpage

\subsection{ICs obtained by gICA}

\begin{figure}[h!]
	\centering
	\begin{subfigure}[b]{0.495\textwidth} 
    	\includegraphics[width=\textwidth]{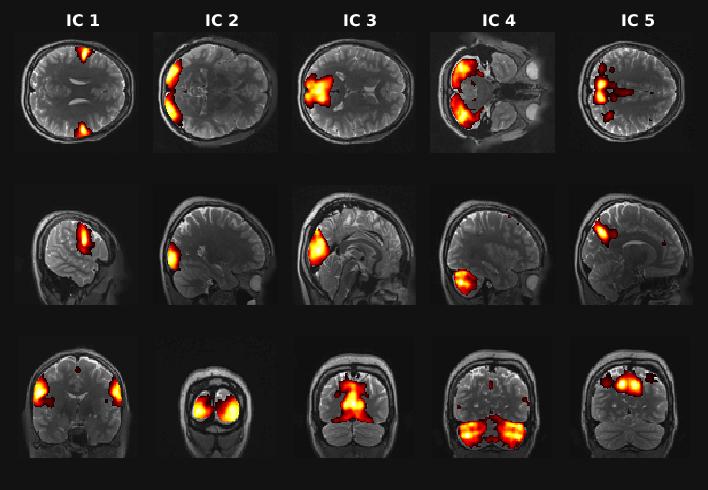}
	\end{subfigure}\hspace{-0.1em}%
	\begin{subfigure}[b]{0.495\textwidth} 
    	\includegraphics[width=\textwidth]{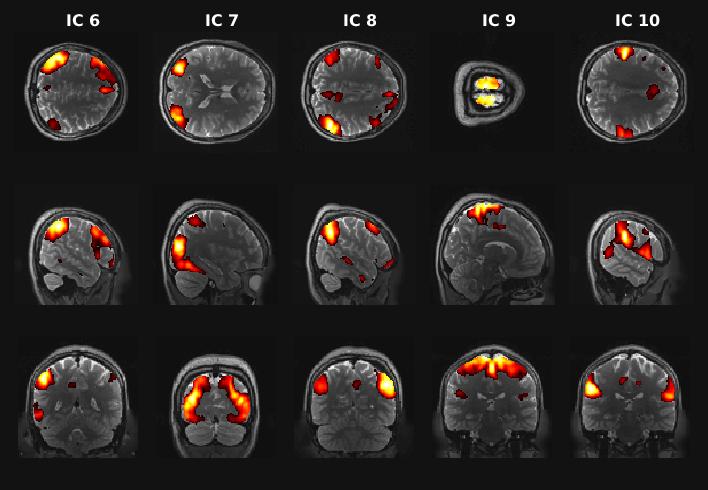}
	\end{subfigure} 

	\begin{subfigure}[b]{0.495\textwidth} 
    	\includegraphics[width=\textwidth]{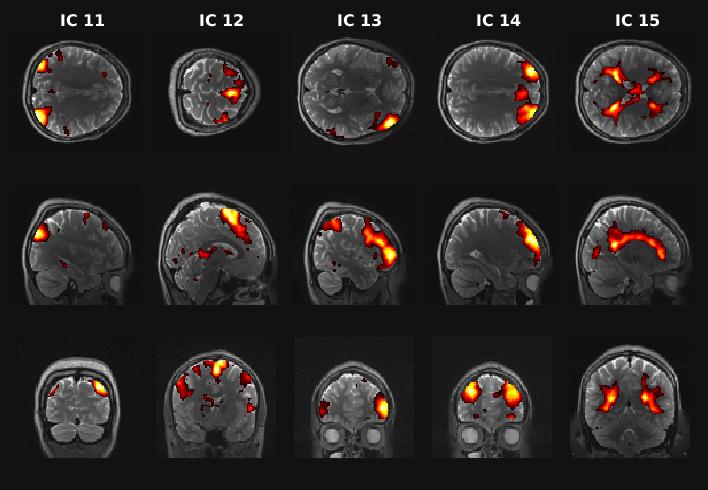}
	\end{subfigure}\hspace{-0.1em}%
	\begin{subfigure}[b]{0.495\textwidth} 
    	\includegraphics[width=\textwidth]{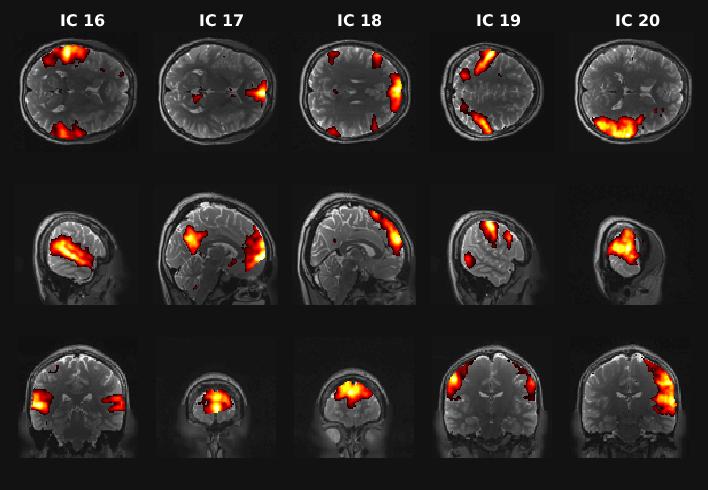}
	\end{subfigure}
	\caption{ICs obtained by gICA. The depicted ICs are computed as mean ICs over subjects. The ICs are presented in the three principal anatomical planes: transverse, sagittal and frontal. The three planes intersect the voxel with highest intensity value in every volumetric image. For visualization purposes, the activations of the networks shown are normalized to zero mean and unit variance, and the intensity of the pictured brain slice $\hat{I}(\mb{r})$ was thresholded by $\hat{I}(\mb{r}) > 2$. The color range is adjusted to the largest intensity value in every pictured slice.}
\end{figure}

\end{document}